\documentclass[allclo]{FBSart}
\usepackage{amsfonts}
\usepackage{amssymb}
\usepackage{epsf}

\title{Scattering of heavy charged particles on hydrogen atoms}

\author{R. Lazauskas\thanks{e-mail: lazauskas@isn.in2p3.fr}
and J. Carbonell\thanks{e-mail: carbonel@isn.in2p3.fr} }

\institute{Institut des  Sciences Nucl\'{e}aires,
          53 av. des Martyrs, 38026 Grenoble, France}

\runningauthor{R. Lazauskas, J. Carbonell}
\runningtitle{Scattering of heavy charged particles on hydrogen atoms}
\begin{document}

\maketitle
\begin{abstract}
The low energy scattering of heavy positively charged particles
on hydrogen atoms (H) are investigated by solving the Faddeev equations in configuration space.
A resonant value of the pH scattering length, $a=750\pm 5$ a.u.,
in the pp antisymmetric state was found.
This large value indicates the existence of a first excited state with
a binding energy  B=1.14$\times$10$^{-9}$ a.u. below the H ground state.
Several resonances for non zero angular momenta states are predicted.
\end{abstract}

\section{Introduction}\label{Intro}

The scattering of heavy charged particles ($X^+$) on 
atoms at
kinetic energies smaller than the inelastic thresholds
is dramatically influenced by the presence of atomic electrons.
Their virtual excitations in presence of the incoming charged particle
result into long range attractive forces which dominate the low energy scattering.
These states were proposed as a possible source of metastability
in the \={p}He system
\cite{CCG_FBS_95} and were found to play a determinant role
in the low energy \={p}H annihilation \cite{VC_PRA_98}.
A full solution of these problems is however made extremely
difficult by the presence of annihilation channels
and only approximate solutions were achieved.
We present in what follows a rigorous solutions for the simplest problem of
scattering on hydrogen atoms.

A simple two-body approximation is first discussed in Section \ref{2B}.
Section \ref{3B} is devoted to the solution of the 3-body (X$^+$e$^-$p$^+$)
problem,  obtained by solving the Faddeev equations in configuration space.
We treat with special care the pH case, for it exhibits the more
interesting properties and constitutes moreover a realistic experimental challenge.
Some final remarks about the present calculations
and plans for future work are given in the conclusion.
We use all along the paper electronic atomic units ($m_e=e^2=\hbar=1$).

%%%%%%%%%%%%%%%%%%%%%%%%%%%%%%%%%%%%%%%%%%%%%%%%%%%%%
\section{Two-body approach}\label{2B}

\newcommand\egal{\mathop{=}}

To get a first qualitative insight into the underlying physics, it is interesting to consider
a simple 2-body X$^+$H problem. The H atom is supposed to be 
point-like
and to interact with the incoming particle via a central potential
\begin{equation}\label{V_pol}
V(r) = \frac{1}{2}\frac{\alpha(r)}{r^4}
\end{equation}
$\alpha(r)$ tends to the H dipole polarizability
($\alpha_d=\frac{9}{2}$) for large values of $r$ and regularizes 
the $\frac{1}{r^4}$ singularity  at $r$=0.
Its precise form is given in \cite{MM_65}.

\begin{figure}[htbp]
\begin{minipage}[t]{64mm}
\begin{center}\epsfxsize=65mm\mbox{\epsffile{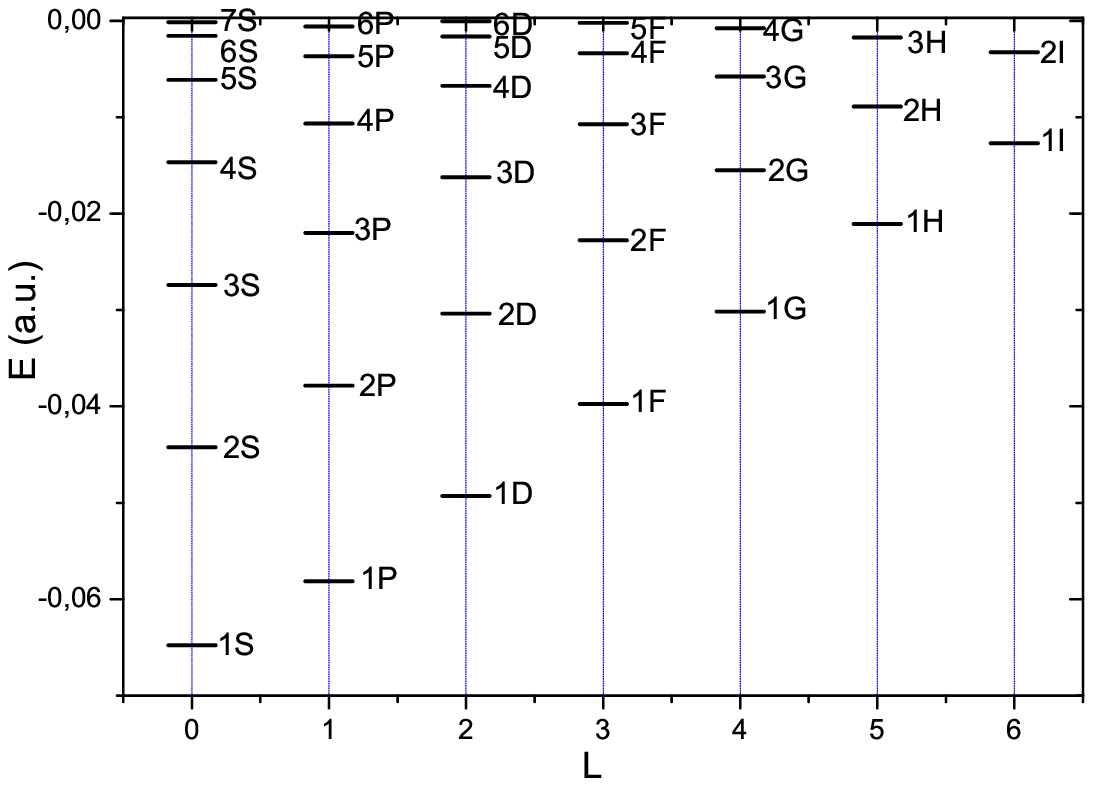}}\end{center}
\caption{Discrete spectrum for $\pi$H in the two body approach}\label{Fig_B_piH}
\end{minipage}
\hspace{0.2cm}
\begin{minipage}[t]{64mm}
\begin{center}\epsfxsize=65mm\mbox{\epsffile{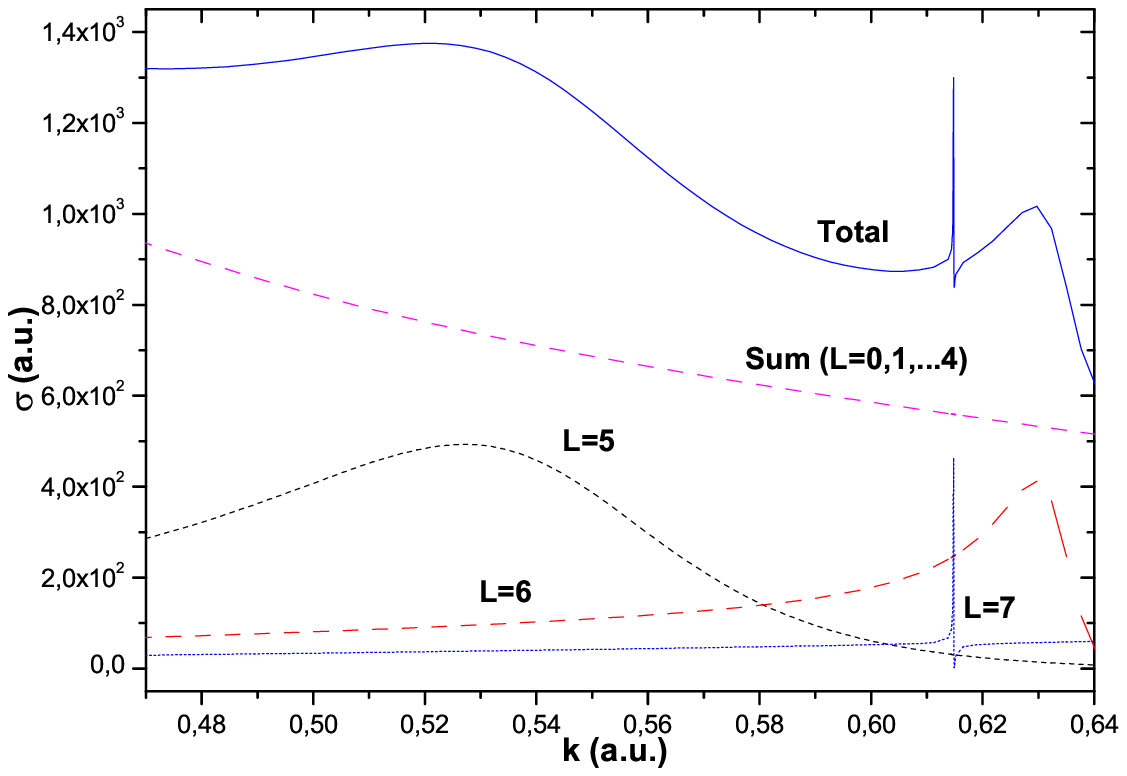}}\end{center}
\caption{Elastic $\mu$H cross section for several partial wave displaying resonant behaviour}\label{Fig_sig_muH}
\end{minipage}
\end{figure}

We have displayed in Fig. \ref{Fig_B_piH} the binding
energies of the $\pi^+$H bound states with interaction (\ref{V_pol}).
One can see a large number of states with angular momentum values up to L=7.
Some of them are very close to the dissociation threshold.
The elastic $\mu^+$H cross section
in the momentum range $k\in[0.48,0.64]$ has been plotted in Fig. \ref{Fig_sig_muH}
with the most relevant partial wave contributions shown separately.
The L=5,6,7 states display a clear resonant behaviour also visible in the total cross section.
A direct calculation of their position and width 
-- using the complex rotation method \cite{CCRM} -- provides the values
E$_{5}$=(7.9-1.4$i$)10$^{-4}$,
E$_{6}$=(1.1-0.25$i$)10$^{-3}$,
E$_{7}$=1.0$\times$10$^{-3}$-1.3$i\times$10$^{-7}$ respectively \cite{RL_DEA_00}.
These resonances concern all high angular momentum states,
involving large centrifugal barriers and relatively large energies.
It's worth noticing however that they can be found even for L=1
and at extremely small energy values.
In the $\pi^+$H case one has e.g. E$_1$=(4.9-1.4$i$)$\times$10$^{-7}$.
The examples shown in Figs. \ref{Fig_B_piH} and \ref{Fig_sig_muH} illustrate well
the kind of physics governed by the polarization forces.
An extensive work as a function of the projectile mass has been done \cite{LC_02}
showing a very rich spectrum
of bound and resonant states, with a complexity increasing  as a function of the projectile mass.

To what extend the results of this simple approach are reliable?
Answering this question was the main motivation of this work.
A definite answer will only appear by letting the electron dynamics in H 
play its
full role, that is by considering the (X$^+$p$^+$e$^-$) three body problem.
And this is the aim of the next section.

%%%%%%%%%%%%%%%%%%%%%%%%%%%%%%%%%%%%%%%%%%%%%%%%%%%%%
\section{Three-body calculations}\label{3B}

The 3-body (X$^+$p$^+$e$^-$) calculations are performed
using Faddeev equations in configuration space.
Three different sets of Jacobi coordinates are involved, defined by
\begin{eqnarray}
\vec{x}_{\alpha}  &=&-\left[\frac{2m_{\beta}m_{\gamma}}{m_{\beta}+m_{\gamma}}\right]^{1/2}
(\vec{r}_{\beta}-\vec{r}_{\gamma}),\label{eq3.1}\\
\vec{y}_{\alpha}  &=&-\left[\frac{2m_{\alpha}(m_{\beta}+m_{\gamma})}
{m_{\alpha}+m_{\beta}+m_{\gamma}}\right]^{1/2}
(\vec{r}_{\alpha}-\frac{\vec{r}_{\beta}m_{\beta}+
m_{\gamma}\vec{r}_{\gamma}}{m_{\beta}+m_{\gamma}})\label{eq3.2}
\end{eqnarray}
\begin{figure}[hbtp]
\begin{center}
\epsfxsize=10cm\epsfysize=3cm\centerline{\epsfbox{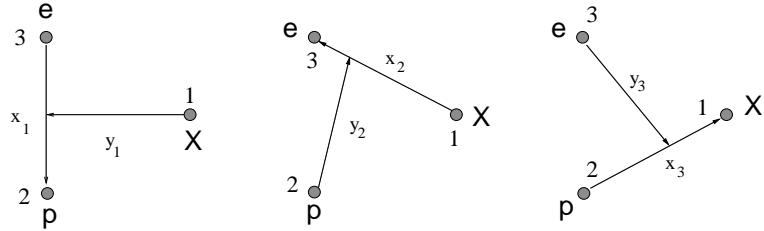}}
\caption{Jacobi coordinates used in X$^+$p$^+$e$^-$ calculations.}\label{Fig 12}
\end{center}
\end{figure}
where $(\alpha\beta\gamma)$ denote cyclic permutations of (123),
and $m_{\alpha}$ the particle masses. We identify 1$\equiv$X$^+$, 2$\equiv$p$^+$, 3$\equiv$e$^-$.
The standard Faddeev  equations read
\begin{equation}\label{FE}
(E-H_{0}-V_{\alpha})\Psi_{\alpha}=V_{\alpha}\sum_{\alpha\neq\beta}\Psi_{\beta},
\end{equation}
where $H_{0}$ is the 3-particle free hamiltonian and
$V_{\alpha}$ the 2-body Coulomb potential for the interacting $(\beta\gamma)$ pair
\begin{equation}
V_{\alpha}(\rm x_{\alpha})=\frac{e_{\beta}e_{\gamma}}
{\mid\vec{r}_{\beta}-\vec{r}_{\gamma}\mid} \qquad e_j=\pm 1\label{eq3.5}
\end{equation}

For projectile masses $m_X< m_p$ we restrict ourselves
 to scattering energies  below
the first rearrangement threshold X$^+$+(p$^+$e$^-$)$\rightarrow$p$^+$+(X$^+$e$^-$). 
In that case, amplitudes $\Psi_2$ and $\Psi_3$ have no asymptotics.

Equations (\ref{FE}) provide satisfactory solutions for bound states
but are not suitable for scattering Coulomb problems.
The reason is that their right hand side does not decrease fast enough
to ensure the decoupling of Faddeev amplitudes in the asymptotic region
and to allow  unambiguous implementation of boundary conditions.
In order to circumvent this problem, Merkuriev \cite{Stas_80} proposed to split
the Coulomb
potential $V$ into two parts by means of some arbitrary cut-off function $\chi$
\begin{eqnarray*}
V(x)     &=& V^s(x,y) +V^l(x,y)  \cr
V^s(x,y) &=& V(x)\chi(x,y) \cr
V^l(x,y) &=& V(x)[1-\chi(x,y)]
\end{eqnarray*}
and to keep in the right hand side of equation (\ref{FE})
only the short range $V_s$ contribution.
One is then let with a system of equivalent equations  
\begin{equation}\label{MFE}
(E-H_{0}-W_{\alpha}-V^s_{\alpha})\Psi_{\alpha}=V^s_{\alpha}\sum_{\alpha\neq\beta}\Psi_{\beta},
\end{equation}
in which $W_{\alpha}$ are some 3-body potential containing the long range parts:
\begin{eqnarray}
W_{\alpha} &=& V^l_{\alpha}+V^l_{\beta}+V^l_{\gamma}
\end{eqnarray}
This approach was found to be very efficient
in calculating the e$^+$Ps and e$^+$H cross sections \cite{KCG_PRA_92,KCG_PRA_95}.
\begin{figure}[htbp]
\begin{center}
\epsfxsize=90mm\epsfysize=80mm\mbox{\epsffile{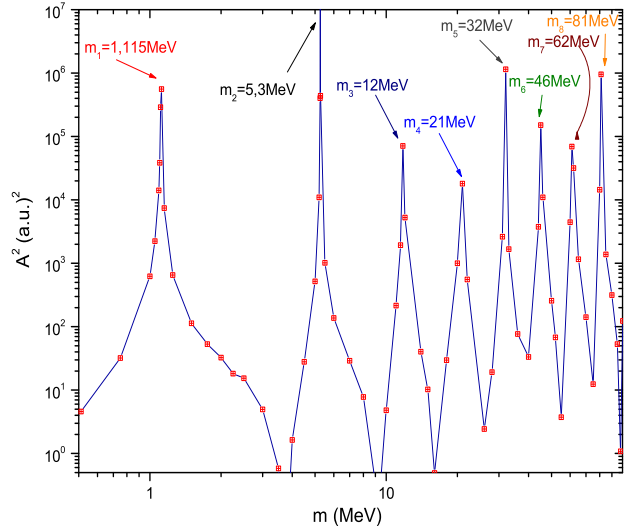}}
\caption{3-body zero energy X$^+$H cross section as a function of 
the projectile mass $m_X$}\label{a2_3c_m}
\end{center}
\end{figure}

Equations (\ref{MFE}) were solved
by expanding $\Psi_i$ in the bipolar harmonics basis
\begin{equation}\label{PW}
\Psi_i(\vec{x}_i,\vec{y}_i)=\sum_{\alpha_i}
{\varphi_{i\alpha_i}(x_i,y_i)\over x_iy_i} \;B_{\alpha_i}^{LM}(\hat{x}_i,\hat{y}_i)
\qquad \alpha_i\equiv\left\{l_{x_i},l_{y_i}\right\}
\end{equation}
and  their components $\varphi_{i\alpha_i}$ in the basis of two-dimensional splines.
In practical calculations we took the cutoff function:
\[ \chi(x,y)=
% \frac{2}{1+\exp{\left[{\left({x\over x_0}\right)^{\nu}\over 1+{y\over y_0}}  \right] }  }
2 \left\{1+\exp{\left[{\left({x\over x_0}\right)^{\nu}\over 1+{y\over y_0}}\right]}\right\}^{-1}\]
Final results are independent of the parameters $x_0,y_0,\nu$ but an appropriate choice
for their values makes the convergence of expansion (\ref{PW}) faster.
The values $x_0=2.0,y_0\approx2\sqrt{m_X},\nu=2.3$ are suitable.

\begin{figure}[htbp]
\begin{minipage}[t]{64mm}
\begin{center}\epsfxsize=64mm\mbox{\epsffile{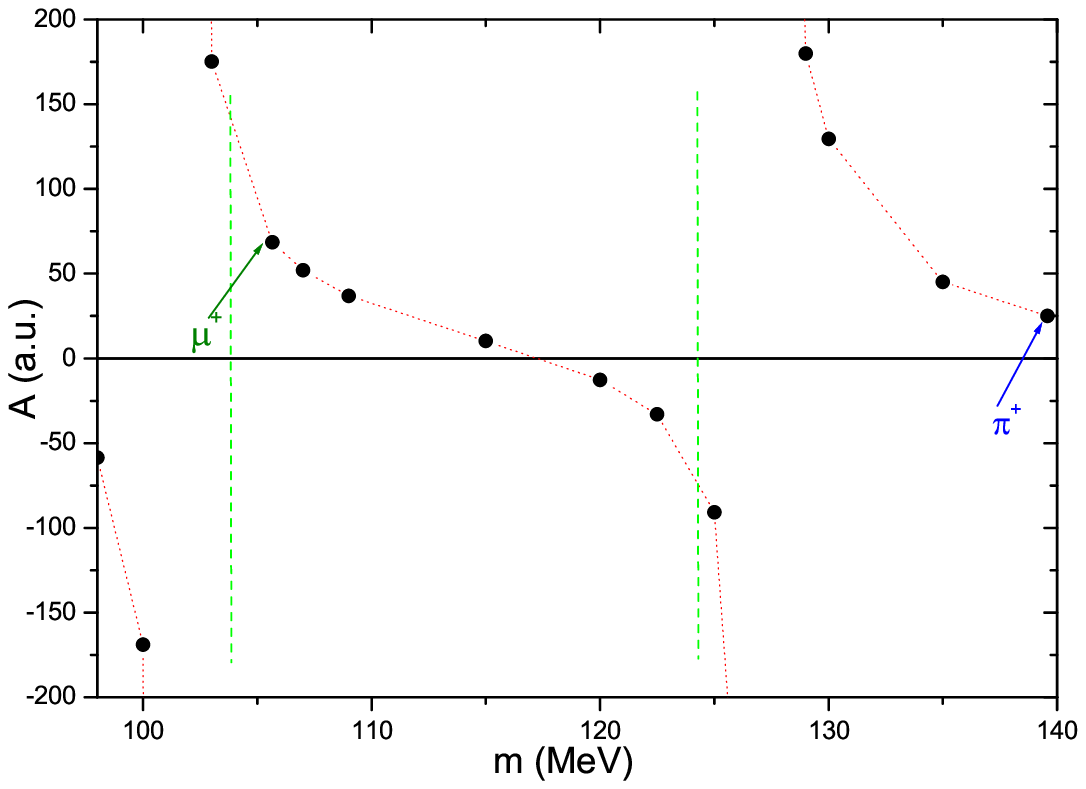}}\end{center}
\caption{Scattering length as a function
of $m_X$, in the region of $\mu$ and $\pi$.}\label{a3c_mupi}
\end{minipage}
\hspace{0.2cm}
\begin{minipage}[t]{64mm}
\begin{center}\epsfxsize=64mm\mbox{\epsffile{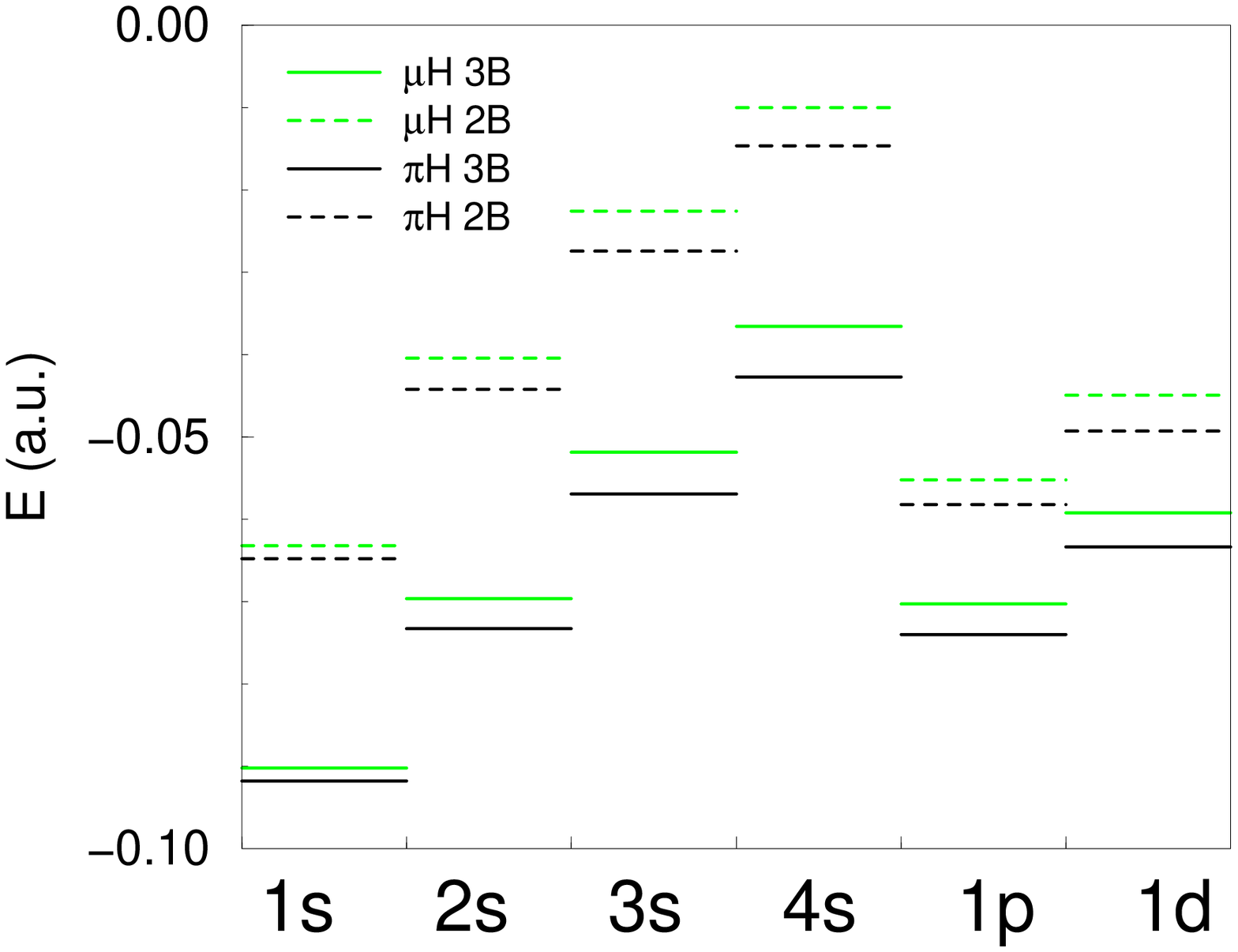}}\end{center}
\caption{Energies for lower  $\pi$H and $\mu$H bound states compared
to  2-body results (\ref{V_pol})}\label{B3c_2c}
\end{minipage}
\end{figure}

Binding energies for the lower 3-body
$\pi^+$H and $\mu^+$H bound states are plotted in Fig. \ref{B3c_2c}.
They are compared to the results of the 2-body approach (\ref{V_pol}).
One can see that, although there is a  qualitative agreement,
2-body energies are systematically underestimated.
These results can be used to improve the short range part of the 2-body potential.

Some interesting features of the 3-body Coulomb system can be learned
from Fig. (\ref{a2_3c_m}), where the zero energy X$^+$H
cross section as a function of the projectile mass $m_X$ is displayed.
Each peak corresponds to the appearence of a new S-wave bound state.
The critical mass values $m_i$ at which they occur,
would enable to generalize the ground state stability
triangle \cite{Martin92} to higher excitations.
A zoom in the region of physical interest, $\mu^+$ and $\pi^+$,
is shown in Fig. \ref{a3c_mupi}.
The calculated values are respectively $a_{\mu H}=69.1$ and $a_{\pi H}=24.4$.
Some care has to be taken in extracting the scattering observables
specially at zero energy, from the asymptotic solution at finite distance.
The long range polarization force makes the
convergence of the observables as a function of the X$^+$-H distance
very slow and requires an appropriate extrapolation procedure \cite{LC_02}.

pH scattering deserves special comments.
The 3-body wave function has to be antisymmetyric with respect
to the p exchange.
This can be realized in two different ways following the proton spin coupling.
For the case when
the two protons spins are antiparallel (singlet) the spatial part of
the wave function is symmetric,
while for the parallel case (triplet) it is antisymmetric.
In the 2-body approach, these two cases give rise to
completely different potentials.

The singlet case has a broad attractive
well which supports a great number of bound states.
They have been calculated since the first days of Quantum Mechanics
and they are presently known with a very high precision
(see e.g.  \cite{TYDB_MP_99,HBGL_EPJD_00} and reference therein).
Our 3-body calculations
cannot reach this kind of accuracy for bound states
but are in good agreement for the lower excitations.
They provide furthermore the first result for
the pH scattering length $a_{s}=-29.3$.
We notice that the zero energy scattering wave function
shows 20 nodes in $y_1$-direction,
indicating the existence of 20 L=0 $\sigma_{g}$
energy levels for $H_{2}^{+}$.

The triplet case -- modeled
by Landau \cite{LL_3} -- is dominated by the Pauli repulsion
between the two protons, overbalanced at $r\sim$10
by the attractive polarization forces.

\begin{figure}[htbp]
\begin{center}
\epsfxsize=120mm\mbox{\epsffile{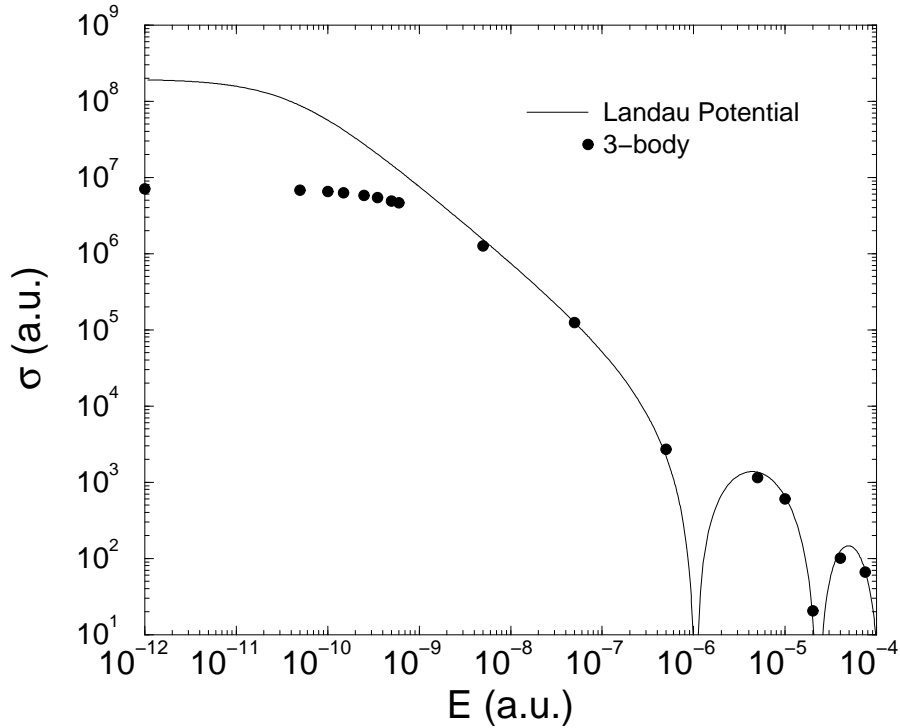}}
\caption{Zero energy pH in the pp triplet state,
compared to the results of Landau potential}\label{sigma_pH_A}
\end{center}
\end{figure}

Our 3-body calculations give a scattering length of $a_{t}=750\pm 5$.
The nodal structure of the Faddeev amplitudes indicates
that such a big value is due to the existence of a
first excited L=0 state with extremely small binding energy.
By using the modified effective range theory \cite{MERT}
we are able to determine its binding energy, which turns to be
B=(1.135$\pm$0.035)$\times$10$^{-9}$  below the H ground state.
To our knowledge, this is
the weakest bond ever predicted, three times smaller
than the $^4$He atomic dimer \cite{Dimer}.

The existence of  L=0 and L=1 ground states is well known and
their binding energies have been very precisely calculated  
\cite{TYDB_MP_99,HBGL_EPJD_00}.
These authors were however not able to conclude about the existence of
a second S-wave bound state.

It is interesting to compare
the three-body calculations with those provided
by the simple Landau two-body potential.
S-wave cross sections are plotted in Fig.  \ref{sigma_pH_A}.
At zero energy, both calculations differ by two orders
of magnitude while at energies E$\sim$10$^{-6}$
they are already in quite a good agreement, despite
the simplicity of the 2-body approach.
This is due to the fact that the effective interaction is highly
repulsive at short distances, what
prevents the incoming proton to penetrate inside the H atom and minimizes
the  effect of the 3-body dynamics.

\begin{figure}[htb]
\begin{minipage}[t]{65mm}
\begin{center}\epsfxsize=65mm\mbox{\epsffile{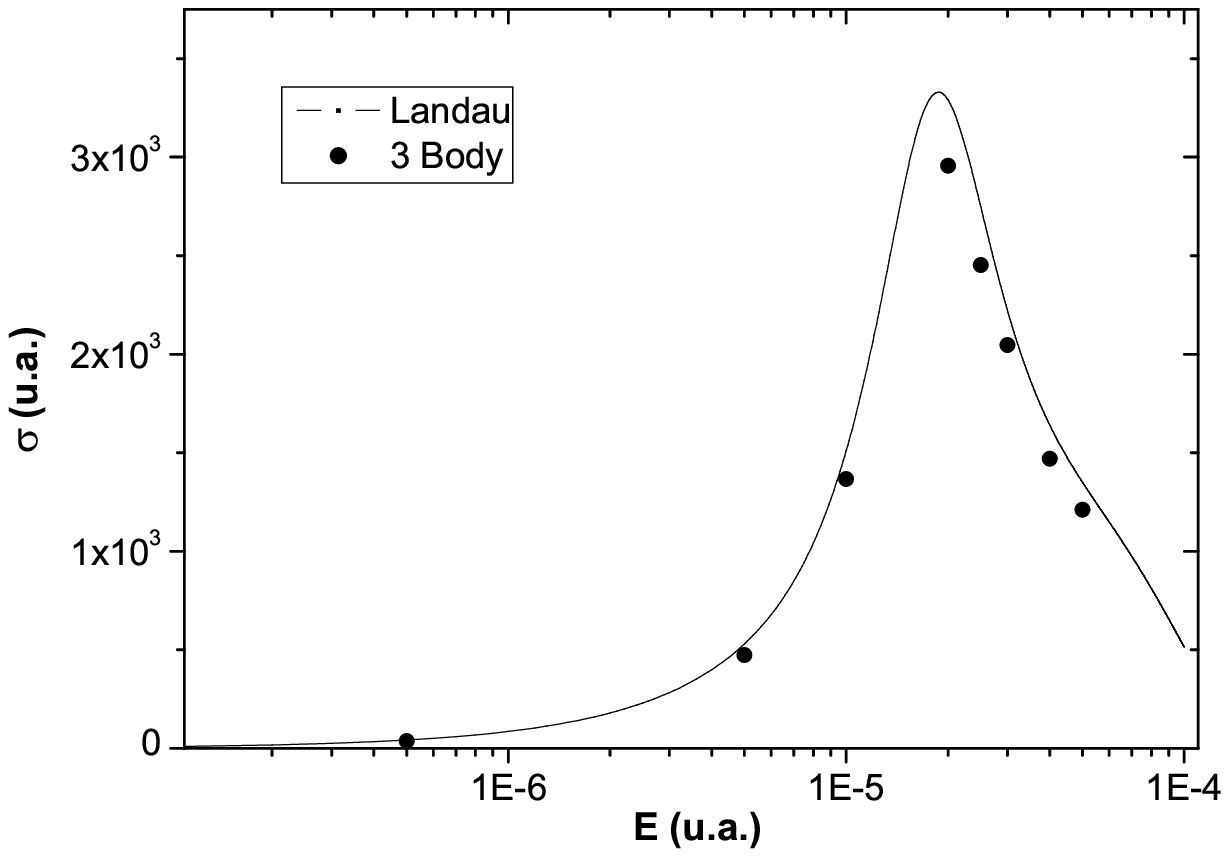}}\end{center}
\caption{pH elastic cross section for L=3 in the  pp spin triplet state}\label{Fig_pH_L3_A}
\end{minipage}
\hspace{0.2cm}
\begin{minipage}[t]{65mm}
\begin{center}\epsfxsize=65mm\mbox{\epsffile{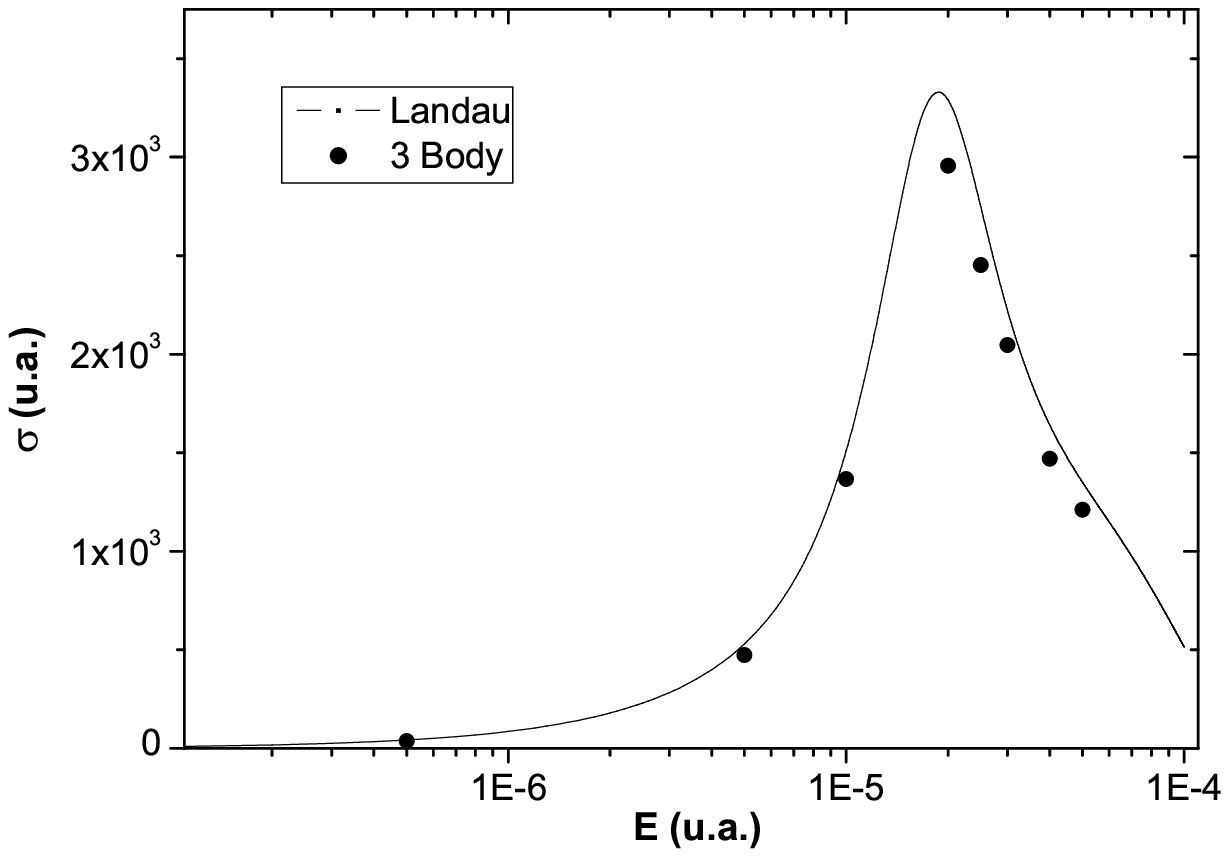}}\end{center}
\caption{pH elastic cross section for L=4 in the  pp spin triplet  state}\label{Fig_pH_L4_A}
\end{minipage}
\end{figure}

The  pH cross sections for higher partial waves have also been calculated.
They exhibit some narrow resonances in several partial waves.
Figs. \ref{Fig_pH_L3_A} and  \ref{Fig_pH_L4_A}
show the elastic cross section for the L=3,4 states. They
are compared to the results of Landau potential.
We can see that their agreement in this energy
region is rather good. The 3-body
resonances are a little bit shifted to the lower energy region
and have smaller
widths. The same effect is seen in the bound state calculations
where the 2-body results are always slightly underbound.
The position and width of these resonances were estimated to
E=(5.13-1.61$i$)$\times 10^{-6}$  for L=3
and E=(1.56-0.94$i$)$\times 10^{-5}$ for L=4.

%%%%%%%%%%%%%%%%%%%%%%%%%%%%%%%%%%%%%%%%%%%%%%%%%%%%%
\section{Conclusions}\label{C}

By solving the Faddeev equations in configuration space,
we have calculated the bound and scattering states for the (X$^+$p$^+$e$^-$)
system with projectile masses $m_X$ larger than the electron.
The long range polarization forces give rise to a rich spectrum of
bound and resonant states with increasing complexity as $m_X$ increases.

Predictions for the physical cases
$X^+$=$\mu^+$, $\pi^+$ ,$K^+$ are obtained. We
found in particular the scattering length values
a$_{\pi^+ H}$=24.4  and a$_{\mu^+ H}$=69.1 a.u.

Of special interest is the pH system in the pp spin-triplet state.
We predict a second S-wave bound state
with binding energy B=1.14$\times$10$^{-9}$ a.u. below the H ground state.
This constitutes the weakest bond ever predicted, 
even smaller than the $^4$He atomic dimer.

The existence of such a nearthreshold state dominates  the low energy pH scattering
and results into a scattering length  of a$_t$=750$\pm$5 a.u.
A low energy proton approaching an H atom will behave like
colliding with a large nanoscopic object.
Several resonances occurring in different partial waves, but visible in the total
cross section, are also predicted.
The experimental confirmation of these results would be very interesting.

The  calculations presented here
are performed using a fully non relativistic dynamics
with Coulomb pair-wise interactions.
In this framework, they are parameter free
with  the only input of particles masses and charges.
The perturbations induced by strong X$^+$-p interactions
or higher electromagnetic corrections have not been included.

In view of the extreme sensibility of the pH results,
it is necessary to quantify both relativistic and strong-interaction effects.
A direct measurement of the pH cross section at very low energy 
seems unlikely.
One can however access the low energy pH continuum 
in the final state of the H$_2^+$
photodissociation cross section.
Work is in progress in these two directions.

\begin{acknowledge}
The authors are sincerely grateful to C. Gignoux
for useful discussions and helpful advises.
The numerical calculations were performed  at CGCV (CEA Grenoble) and  IDRIS
(CNRS). We thank the staff members  of these organizations for their constant support.
\end{acknowledge}


\begin{thebibliography}{9}
%\bibitem{H} Hearing, W., et al.: Nuovo Cim. {\bf A35}, 345 (1979)
\bibitem{CCG_FBS_95} J. Carbonell, J., F. Ciesielski, F., Gignoux, C.: Few-Body Systems Suppl. {\bf 8}, 428 (1995)
\bibitem{VC_PRA_98}  Voronin, A., Carbonell, J.: Phys. Rev. {\bf A57} 4335 (1998).
\bibitem{MM_65} Mott, N.F., Massey H.S.W.: The Theory of Atomic Collisions.
                \\ Oxford Science Publications $3^\textrm{rd}$ Edition 1965
\bibitem{CCRM}  Moiseyev, N.:  Phys. Rep. {\bf 302}  211 (1998)
\bibitem{RL_DEA_00} R. Lazauskas, R.: Rapport de Stage DEA, ISN-UJF (2000)
\bibitem{LC_02} R. Lazauskas, R., Carbonell, J.: to be published
\bibitem{Stas_80} Merkuriev, S.P. : Ann. Phys. {\bf 130}, 395 (1980)
\bibitem{KCG_PRA_92}  A. A. Kvitsinsky, J. Carbonell, C. Gignoux:  Phys Rev. {\bf A46}  (1992)  1310
\bibitem{KCG_PRA_95}  A. A. Kvitsinsky, J. Carbonell, C. Gignoux:  Phys. Rev.  {\bf A51} (1995) 2997
\bibitem{Martin92}    A.~Martin, J.-M.~Richard and T.T.~Wu, Phys.\ Rev.\ {\bf A46} (1992) 3697
\bibitem{TYDB_MP_99}  Taylor, J.M., Yan, Z., Dalgarno, A., Babb, J.F.:  Molec. Phys. {\bf 97}, 25 (1999)
\bibitem{HBGL_EPJD_00}  Hilico, L., Billy, N., Gremaud, B., Delande, D.: Eur. Phys. J. {\bf D12}, 449 (2000)
\bibitem{Dimer} F. Luo, G. McBane, G. Kim, C. Giese, W. Gentry, J. Chem. Phys. {\bf 98} 3564 (1993)
\bibitem{LL_3} Landau, L., Lifshits, E.: Mecanique Quantique, Ed. Mir Moscou  1975,
\bibitem{MERT}  O'Malley T.F., Spruch L., Rosenberg L., J.:  Phys. Rev. {\bf 125}  491 (1961)
\end{thebibliography}
\end{document}